\documentclass{IEEEtran}
\usepackage{authblk}
\usepackage{keyval}

\usepackage[utf8]{inputenc} 
\usepackage[T1]{fontenc}    
\usepackage{hyperref}       
\usepackage{url}            
\usepackage{booktabs}       
\usepackage{amsfonts}       
\usepackage{nicefrac}       
\usepackage{subcaption}     
\usepackage{nicefrac}       
\usepackage{subcaption}     
\usepackage{microtype}      
\usepackage{lipsum}		

\usepackage{graphicx}
\usepackage{natbib}
\usepackage{doi}
\usepackage{flushend}

\title{Towards fairer public transit: Real-time tensor-based multimodal fare evasion and fraud
detection}
\date{September 9, 1985}	
\author{Peter Wauyo, Dalia Bwiza\\
Carnegie Mellon University Africa\\ Kigali, Rwanda\\
\texttt{[pwauyo, bdalia]@andrew.cmu.edu}\\
Alain Murara\\
Rwanda Utilities Regulatory Authority\\
\texttt{alain.murara@rura.rw}\\
\href{https://orcid.org/0000-0003-0985-3002}{\includegraphics[scale=0.06]{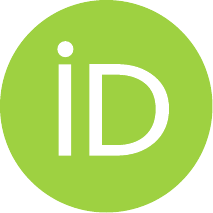}\hspace{1mm}Edwin Mugume},  
	\href{https://orcid.org/0000-0002-2451-8897}{\includegraphics[scale=0.06]{orcid.pdf}\hspace{1mm}Eric Umuhoza} \\
	Carnegie Mellon University Africa\\ Kigali, Rwanda\\
	\texttt{[emugume, eumuhoza]@andrew.cmu.edu} \\
}
\begin{document}
\maketitle

\begin{abstract}
This research introduces a multimodal system designed to detect fraud and fare evasion in public transportation by analyzing closed circuit television (CCTV) and audio data. The proposed solution uses the Vision Transformer for Video (ViViT) model for video feature extraction and the Audio Spectrogram Transformer (AST) for audio analysis. The system implements a Tensor Fusion Network (TFN) architecture that explicitly models unimodal and bimodal interactions through a 2-fold Cartesian product. This advanced fusion technique captures complex cross-modal dynamics between visual behaviors (e.g., tailgating, unauthorized access) and audio cues (e.g., fare transaction sounds). The system was trained and tested on a custom dataset, achieving an accuracy of 89.5\%, precision of 87.2\%, and recall of 84.0\% in detecting fraudulent activities, significantly outperforming early fusion baselines and exceeding the 75\% recall rates typically reported in state-of-the-art transportation fraud detection systems. Our ablation studies demonstrate that the tensor fusion approach provides a 7.0\% improvement in the F1 score and an 8.8\% boost in recall compared to traditional concatenation methods. The solution supports real-time detection, enabling public transport operators to reduce revenue loss, improve passenger safety, and ensure operational compliance.
\end{abstract}


\section{Introduction}

Public transportation in Rwanda plays a crucial role in daily mobility, particularly in urban centers such as Kigali, where thousands of passengers rely on buses as their primary mode of transport. However, evasion of the fare and fraudulent activities remain persistent challenges, leading to significant revenue losses and operational inefficiencies. These issues not only affect transport companies, but also strain government efforts to maintain a reliable and sustainable public transport system\cite{currie2017empirical}.

Computer vision and machine learning technologies offer promising solutions to these challenges. Using closed-circuit television (CCTV) footage, automated systems can be developed to monitor, detect, and prevent fraudulent activities in real time.

The London Underground has successfully implemented a computer vision system to detect fare evasion, resulting in significant reductions in revenue loss by analyzing CCTV footage to identify fraudulent activities \cite{noauthor_london_nodate}. Similarly, the Hong Kong mass transit railway system has integrated CCTV footage with fare transaction data, allowing more accurate and timely fraud detection, which has contributed to improved fare compliance and reduced revenue losses \cite{zhou_intermodal}.
Building on these advancements, our research applies similar techniques to address fare fraud and evasion in the specific context of public transport in Rwanda.



\begin{figure*}[t!]
    \centering
    \begin{subfigure}[t]{0.5\textwidth}
        \centering
       \includegraphics[width=\linewidth]{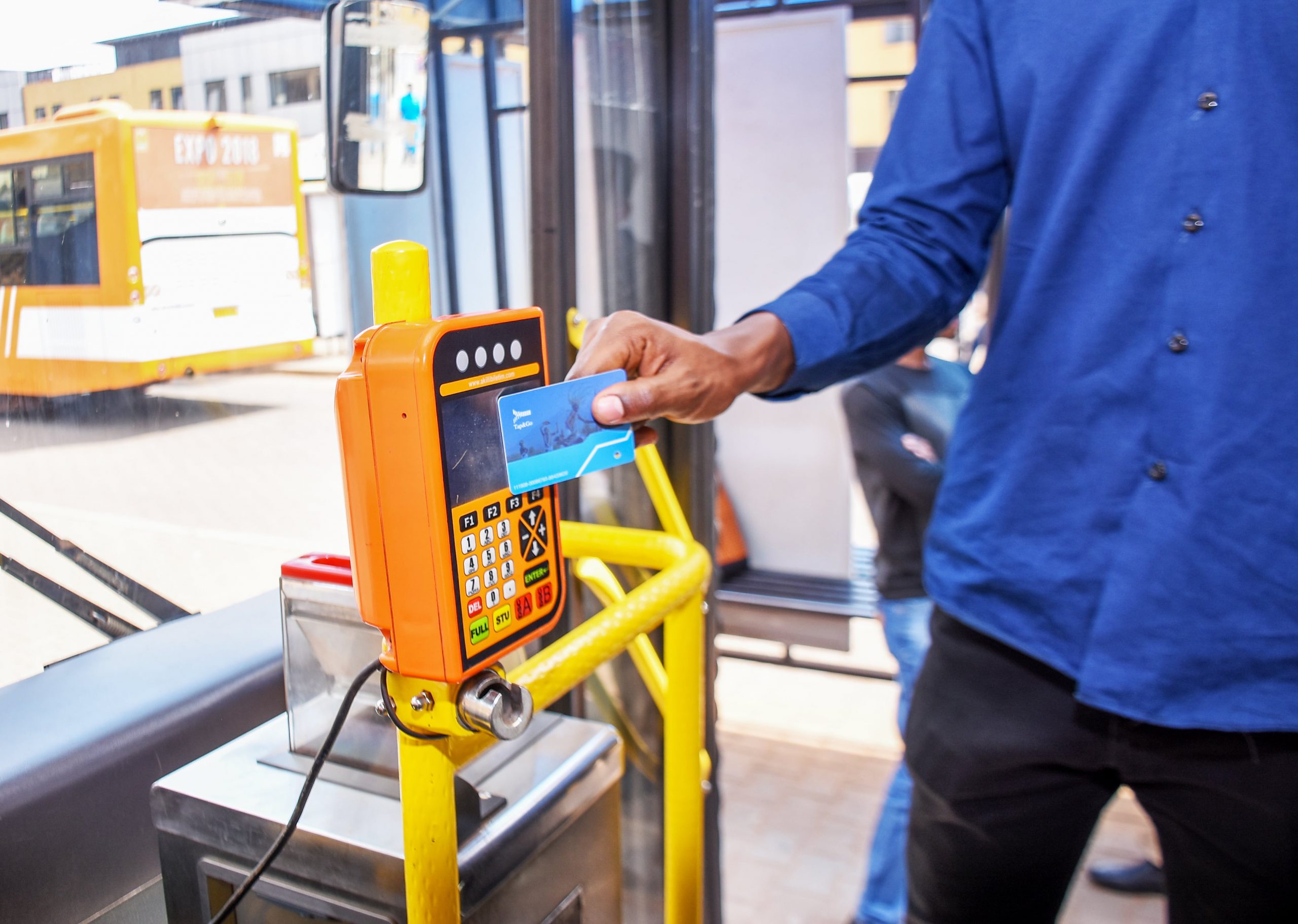}
        \caption{Payment validator}
    \end{subfigure}%
    ~ 
    \begin{subfigure}[t]{0.5\textwidth}
        \centering
       \includegraphics[width=\linewidth]{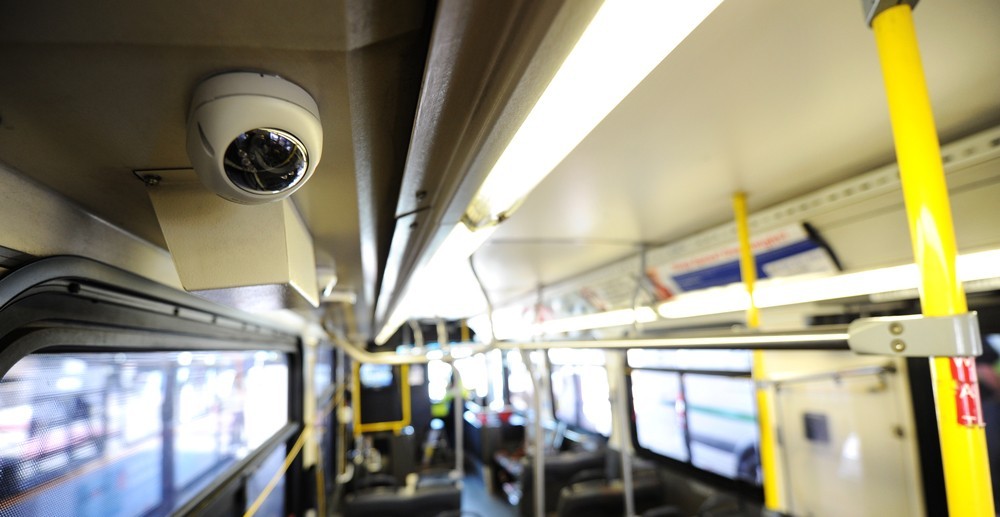}
        \caption{CCTV cameras installed in public buses.}
    \end{subfigure}
     ~ 
    \begin{subfigure}[t]{0.5\textwidth}
        \centering
       \includegraphics[width=\linewidth]{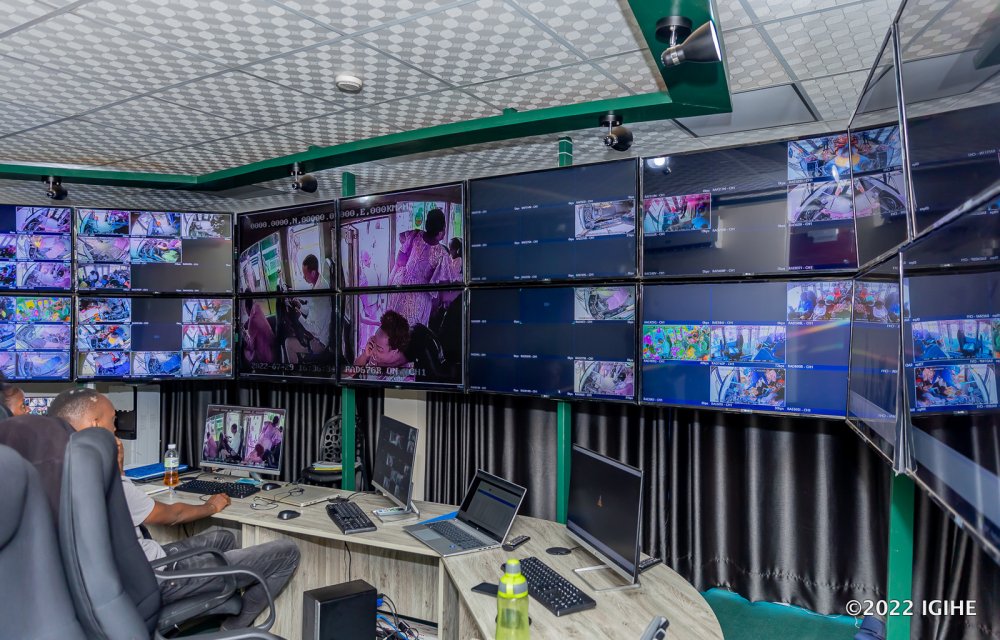}
        \caption{A typical control room}
    \end{subfigure}
    \caption{Current system for detecting fare evasion in Rwanda:
(a) Public buses are equipped with smart card readers that function as payment validators; 
(b) CCTV cameras installed near the validators capture video footage of passengers as they board and alight; and (c) This footage is streamed in real time to a control room, where personnel monitor multiple feeds simultaneously in an attempt to identify passengers who fail to tap their cards.}
    \label{fig:tap&go}
\end{figure*}

This paper presents a novel multimodal system for detecting fraud and fare evasion in public transportation by analyzing both CCTV video and audio data. The key contributions of this research are threefold:
\begin{enumerate}
  
 \item \emph{A multi-modal fusion architecture for fraud detection.} Our study adapts the Tensor Fusion Network (TFN) architecture, originally developed by Zadeh et al.\cite{zadeh2017tensor} for multimodal sentiment analysis, and extends it to the domain of fraud detection. This extended architecture enables the system to capture complex cross-modal dynamics--such as the interplay between visual behaviors (e.g. tailgating, unauthorized access) and audio cues (e.g., fare transaction sounds)--enabling the detection of subtle fraud indicators that are often overlooked by conventional fusion methods.\\

 \item \emph{Integration of state-of-the-Art feature extraction Models.}  The proposed solution incorporates the Vision Transformer for Video (ViViT) \cite{arnab2021vivit} model to extract high-level spatio-temporal features from CCTV footage and the Audio Spectrogram Transformer (AST) \cite{gong2021ast} to extract discriminative features from the audio input. These models operate synergistically within the TFN framework, enabling robust processing of real-time surveillance data from public transport environments.\\

 \item \emph{Empirical validation and significant performance improvements.} 
   The proposed system was trained and evaluated on a custom data set adapted to transportation fraud scenarios. It achieved an accuracy of 89. 5\%, a precision of 87. 2\%, and a recall of 84. 0\%, substantially outperforming the early fusion baselines and exceeding the 75\% recall typically reported in state-of-the-art fraud detection systems. Ablation studies further demonstrate that the TFN architecture provides a 7. 0\% gain in the F1 score and an 8.8\% increase in recall over traditional feature concatenation methods.
\end{enumerate}
In addition to the aforementioned scientific contributions, the proposed system offers real-time detection capabilities that can help public transport authorities reduce revenue losses, ensure operational compliance, and improve passenger safety. If adopted, it has the potential to deliver direct economic benefits, particularly in emerging markets.

The remainder of this paper is structured as follows: Section \ref{background} discusses current methods for detecting fare evasion in Rwanda and highlights their limitations. Section \ref{related} reviews related work in automated fraud detection and multimodal analysis. Section \ref{approach} details the proposed methodology, including data collection, feature extraction, and the tensor fusion architecture. Section \ref{res-disc} presents experimental results and analysis, including comparative performance and ablation studies. Section \ref{concl} concludes the paper and outlines directions for future research.

\section{Background: Current methods for detecting fare evasion in Rwanda}
Fare evasion continues to pose a serious challenge within the Rwandan public transportation system, particularly across bus networks operating under the smart card fare collection system. Detection efforts currently rely on manual surveillance of closed-circuit television (CCTV) footage, monitored from a centralized control center managed by JALI (Joint Agency for Local Integrated Transport), the national body responsible for ensuring the compliance of the fare and overseeing the smart card infrastructure \cite{jali2025}.

To facilitate compliance, all public buses are fitted with CCTV cameras that capture video footage of passengers boarding and alighting, as shown in Figure \ref{fig:tap&go}. This footage is streamed in real time to a control room, where personnel monitor multiple feeds simultaneously in an attempt to identify passengers who do not tap their cards. However, this system lacks automation: It relies entirely on the vigilance of human operators to visually detect incidents of fare evasion. As a result, it suffers from several critical limitations, including subjectivity, human error, and limited scalability.

The control room operates 24/7, with four employees working alternating shifts. Each operator is typically responsible for monitoring footage from 10 to 15 different buses at once. Despite their best efforts, several challenges compromise the effectiveness of this manual surveillance approach.

\begin{itemize}
    \item \textbf{Cognitive fatigue}: Prolonged monitoring of multiple video feeds leads to mental fatigue, reducing attention span and detection accuracy. Research shows that human attention deteriorates significantly after just 20 minutes of continuous video surveillance \cite{sulman2008effective}.
    
    \item \textbf{Delayed enforcement}: Once an incident is identified, the employee must manually report it to the relevant authorities or bus operators. This delay often results in missed opportunities to intervene, as the offender may already have exited the bus.
    
    \item \textbf{Limited detection rates}: The combination of screen overload and potential distractions leads to a high rate of missed violations. In similar transit systems, only about 40\% of the fare evasion incidents are successfully detected through manual monitoring \cite{sulman2008effective}.
\end{itemize}

These limitations underscore the need for automated and intelligent fare monitoring systems capable of operating in real time, reducing the dependence on human operators, and significantly improving the detection accuracy.

\label{background}

\section{Related Work}

Various studies have explored the application of computer vision and machine learning techniques to enhance security measures and ensure fare compliance. This section examines existing research on using CCTV data and related technologies for fraud detection in public transport systems.

Traditional methods for detecting fraud and fare evasion in public transportation systems rely heavily on human operators and manual processes. However, these approaches are labor intensive, time consuming, and prone to human errors \cite{bieler_survey_2022}. Random ticket inspections by inspectors serve as a deterrent, but are not comprehensive and may miss many fare evasion incidents \cite{barabino_evaluating_2023}.

Automated fare collection systems, such as contactless card readers, have been implemented to reduce fare evasion. However, these systems primarily address fare evasion at the point of entry and do not effectively handle physical breaches such as tailgating, where an individual follows another through the fare gate without paying \cite{du_detecting_2019}.


Computer vision techniques have shown promise in detecting fare evasion and other fraudulent activities. An effective application is the detection of tailgating. Tuomola et al.  \cite{tuomola_applying_2019} developed a system using computer vision algorithms to detect tailgating incidents by analyzing the flow of passengers through the fare gates. Their approach used background subtraction and object tracking to identify tailgating instances. 

Computer vision-based behavior analysis is another important area of research. Kim et al. \cite{kim_evaluation_2021} demonstrated how the analysis of passenger behavior patterns using CCTV footage can help identify anomalies indicative of evasion of charges or fraudulent activities. Their system used techniques such as optical flow analysis to track and analyze movement patterns within the transport system.

Machine learning models and deep learning techniques, particularly Convolutional Neural Networks (CNNs) and Recurrent Neural Networks (RNNs), have been extensively researched for their high accuracy in image and video analysis tasks \cite{marchetti_study_2023}. CNNs are effective in extracting spatial features from CCTV footage, making them suitable for identifying visual patterns associated with fraudulent activities. RNNs, on the other hand, are adept at handling temporal data, enabling the analysis of sequences in video frames to detect irregular behavior over time. Davis et al. \cite{davis_framework_2020} developed a machine learning model that was trained on historical data to identify deviations from normal passenger behavior, effectively identifying potential fraud cases. Their model used unsupervised learning techniques to detect outliers in the data, which were then reviewed for possible fraudulent activity.

Recent advances in multimodal learning have explored various approaches beyond simple feature concatenation. In particular, Zadeh et al. \cite{zadeh2017tensor} introduced the TFN for multimodal sentiment analysis, which explicitly models interactions between different modalities through a three-fold Cartesian product. Their work demonstrated significant improvements over traditional fusion approaches by capturing complex intermodal dynamics. Although this approach was initially applied to sentiment analysis in conversational videos, our work adapts and extends it to the domain of fraud detection in public transportation.

Integrating CCTV footage with other data sources improves the accuracy and effectiveness of fraud detection systems. Fare transaction records provide a valuable data source that can be correlated with visual anomalies detected in CCTV footage. Shpyrko et al. \cite{shpyrko_fraud_2019} demonstrated how combining these data sources allowed for more robust fraud detection by verifying whether the visual entry of passengers matched the recorded fare transactions.
 
 Passenger profiles and travel history further enrich the data set used for fraud detection. Du et al. \cite{du_detecting_2019} showed that incorporating passenger profiles, including travel frequency and patterns, into machine learning models improved the accuracy of detecting fraudulent activities. This integration allowed the system to account for legitimate variations in passenger behavior.
 
Other works have focused on the possible privacy and security challenges of such approaches in public transportation systems. The use of CCTV footage in public transport raises significant privacy concerns. PrivComBermuda \cite{privcombermuda_cctv_2023} emphasized the need to adhere to data protection regulations, such as the General Data Protection Regulation (GDPR), to ensure that passenger privacy is not compromised. Ethical considerations must also be addressed, balancing the need for security with individual privacy rights \cite{zimmer2005surveillance}.

Furthermore, developing models that perform reliably under diverse and real-world conditions requires robust training and testing methodologies \cite{dou_enhancing_2020}. However, obtaining high-quality data to train these models is a significant technical challenge. 
\begin{figure*}[!h]
    \centering
    \includegraphics[width=\textwidth]{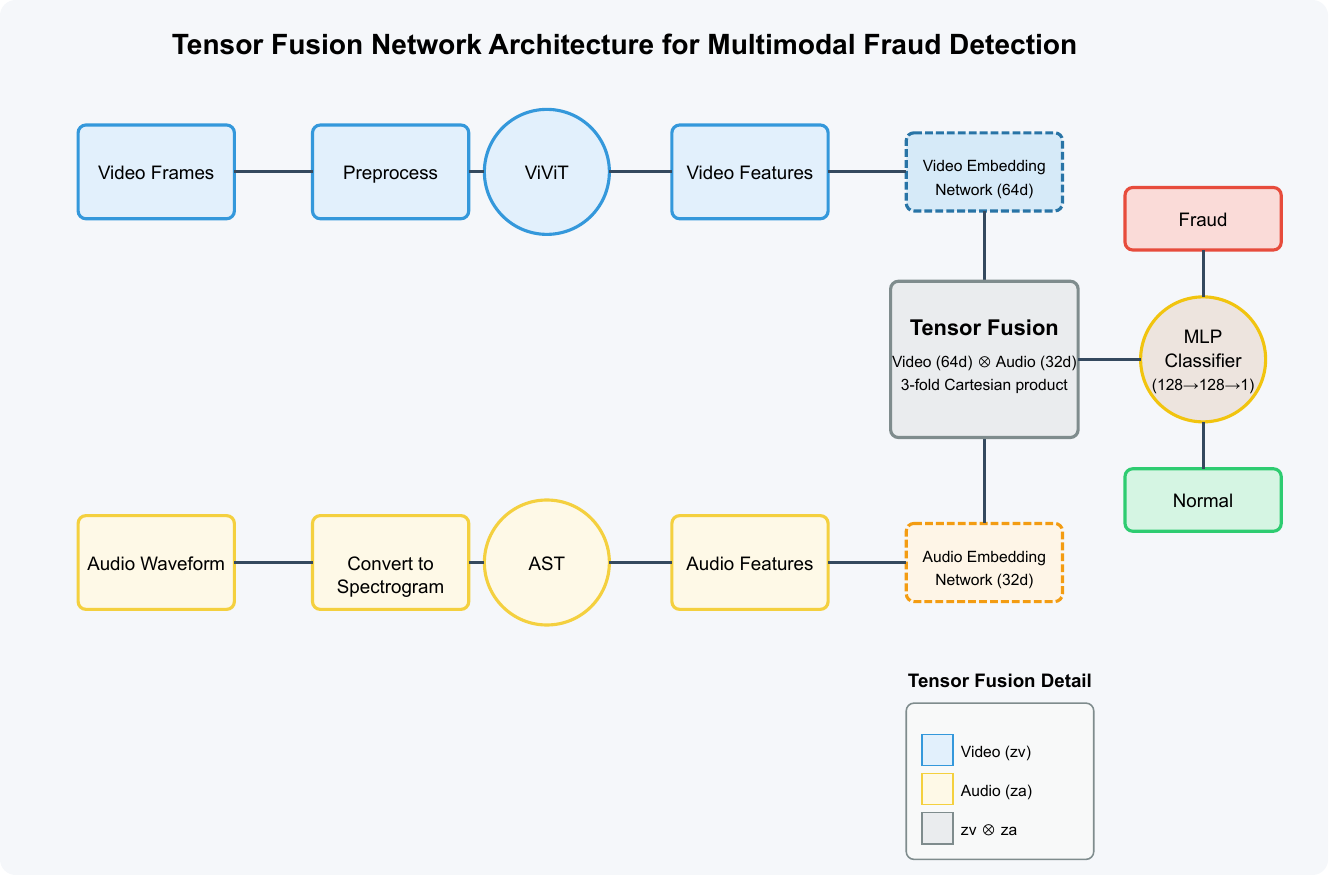}
  \caption{TFN architecture for multimodal fraud detection in Public Transportation Systems.}
    \label{fig:model-architecture}
\end{figure*}

\label{related}

\section{Methodology}
This section outlines the methodology used to detect ticket fraud and evasion in public
transport. The approach combines video and audio analysis to
identify suspicious activities such as bypassing the electronic payment system, making cash
payments to the conductor, and pretending to use a card.

\subsection{System overview}
As shown in Figure \ref{fig:model-architecture}, the proposed model integrates visual and auditory data to detect fraudulent activities in public transportation systems. This multimodal approach takes advantage of the strengths of two state-of-the-art models: ViViT for video data and AST for audio data. Unlike traditional approaches that use simple concatenation for multimodal fusion, our architecture implements the Tensor Fusion Network (TFN) that explicitly models cross-modal interactions through a 2-fold Cartesian product operation.

\subsection{Data collection and preprocessing}
The data used in this study are video footage from various public buses, with a focus on the entrance area where both the payment system and the conductor are visible. In addition, audio recordings were captured to document the distinct sounds produced by the payment system, allowing differentiation between successful and failed transactions.\\
The data were preprocessed as follows:


\begin{enumerate}
    \item \emph{Frame extraction at regular intervals.} The video data was first processed by extracting frames at regular intervals, followed by normalization and resizing to fit the input dimensions required by ViViT. We chose ViViT because of its ability to capture spatio-temporal features \cite{arnab2021vivit}, which are critical in detecting nuanced fraudulent activities.
        
    \item \emph{Data augmentation--random horizontal flipping.} The frames were randomly flipped horizontally to introduce variability. The frames were then cropped to focus on the central part of the frames to focus on the area of interest. The frames were then randomly cropped to create multiple variations. 
        
    \item The audio data was pre-processed by converting it to spectrograms, which were then fed into the Audio Spectrogram Transformer (AST) for feature extraction. AST was selected for its robustness in handling various audio characteristics \cite{gong2021ast}, making it ideal for detecting anomalies that might indicate fraudulent behavior.
        
    \item The audio recordings were segmented to isolate the sounds corresponding to each transaction, and the audio signals were converted to spectrograms for easier analysis.
\end{enumerate}

\subsection{Feature extraction}
Feature extraction forms the foundation of our multimodal fraud detection system, transforming raw video and audio inputs into discriminative representations suitable for analysis. Our approach leverages two state-of-the-art transformer architectures: Vision Transformer for Video (ViViT) to capture spatio-temporal patterns in passenger behavior, and Audio Spectrogram Transformer (AST) to analyze transaction-related acoustic signatures. These models were specifically chosen for their ability to process sequential data and capture long-range dependencies critical for identifying subtle fraud indicators. The following subsections detail how each modality is processed to extract meaningful features that serve as inputs to our tensor fusion network.

\subsubsection{Video feature extraction using ViViT}
The ViViT model is used to extract high-level spatio-temporal features from video footage:
\begin{itemize}
    \item The input consists of frames extracted from the video data, which are resized and normalized.
    \item Each video frame is divided into nonoverlapping patches. These patches are flattened and embedded into a larger space.
    \item The embedded patches are processed through a series of transformer layers, which capture spatial and temporal dependencies within the video data.
    \item The output of the transformer layers is pooled to generate a fixed-length feature vector that represents the video content.
\end{itemize}

\subsubsection{Audio feature extraction using AST}
The AST model is used to extract features from audio data corresponding to the video:
\begin{itemize}
    \item The audio is first converted into a mel-spectrogram, a 2D representation of the audio frequency content over time.
    \item The Mel spectrogram is divided into patches, which are flattened and embedded.
    \item These embedded patches are passed through transformer layers to capture audio patterns related to fraudulent activity.
    \item The features are pooled over time, producing a fixed-length feature vector summarizing the audio information.
\end{itemize}

\subsection{Modality-specific embedding networks}
Before fusion, each modality's features are processed through dedicated embedding networks:
\begin{itemize}
    \item \emph{Video embedding network:} Takes the output representation of ViViT (CLS token).
    Processes through two fully connected layers $(768\rightarrow 128\rightarrow 64)$ with ReLU activations.
    Outputs a $64$-dimensional embedding that captures essential video features.
    \item \emph{Audio embedding network:} Takes the output representation from AST (CLS token).
    Processes through two fully connected layers $(768\rightarrow 128\rightarrow 32)$ with ReLU activations.
    Outputs a $32$-dimensional embedding that captures essential audio features.
\end{itemize}
These embedding networks, detailed in Table \ref{tab:embedding_networks}, serve three key purposes: reducing the dimensionality of transformer output, extracting task-specific features relevant to fraud detection, and transforming features into a compatible representation space for the tensor fusion operation.

\begin{table}[t]
\caption{Configuration of Modality-Specific Embedding Networks}
\label{tab:embedding_networks}
\centering
\begin{tabular}{|l|l|l|}
    \hline
    \textbf{Parameter} & \textbf{Video Network} & \textbf{Audio Network} \\
    \hline
    Input dimension & 768 & 768 \\
    \hline
    Hidden layer size & 128 & 128 \\
    \hline
    Output dimension & 64 & 32 \\
    \hline
    Activation & ReLU & ReLU \\
    \hline
\end{tabular}
\end{table}

\subsection{Tensor fusion layer}

The tensor-fusion layer explicitly models 3 types of multimodal dynamics.
Let $\mathbf{z}_v$ be the 64-dimensional video embedding vector, and $\mathbf{z}_a$ be the 32-dimensional audio embedding vector;

\begin{enumerate}
    \item \emph{Unimodal dynamics:} Preserves unimodal information by appending a constant $'1'$ dimension to each modality embedding, creating extended embeddings $[z_v;1]$ and $[z_a;1]$.
     
   \item \emph{Bimodal dynamics:} Captures cross-modal interactions between video and audio through an outer product operation that creates video-audio interactions: $z_v \otimes z_a$.

    \item  \emph{Trimodal dynamics:} Although our current implementation focuses primarily on two modalities (video and audio), the architecture is extensible to incorporate additional modalities such as textual transaction data in future iterations.
\end{enumerate}
The mathematical formulation of the tensor fusion operation is expressed as:
    \begin{equation}
        z_{fusion} = [z_v;1] \otimes [z_a;1],
    \end{equation}
where $\otimes$ denotes the outer product operation, resulting in a tensor that captures all possible multiplicative interactions between modalities. This creates a tensor of shape $65 × 33 = 2,145$ dimensions.

\begin{figure}[t]
    \centering
    \includegraphics[width=\linewidth]{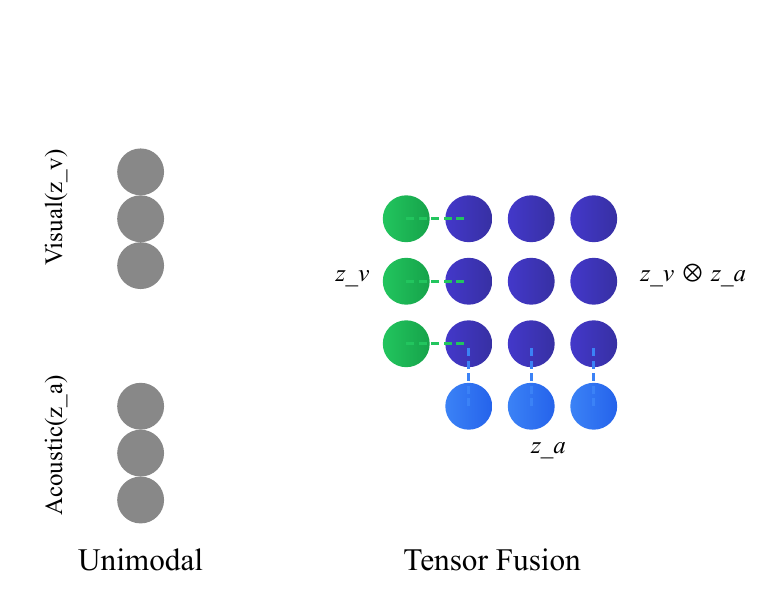}
    \caption{Multimodal fusion architecture for video and audio fraud detection.}
    \label{fig:tensor_fusion}
\end{figure}

Figure \ref{fig:tensor_fusion} illustrates the interaction between unimodal representations and tensor fusion. Individual video $(z_v)$ and audio $(z_a)$ modalities are represented in green and blue, respectively, while their bimodal fusion $(z_v \otimes z_a)$ is shown as a $3\times3$ grid of deep blue nodes in the center. This visualization demonstrates how the tensor fusion combines the separate modalities into a unified representation that captures cross-modal interactions.
The fusion tensor contains seven distinct semantic regions:

\begin{enumerate}
    \item Unimodal video $(z_v)$
    \item Unimodal audio $(z_a)$
    \item Bimodal interaction $(z_v \otimes z_a)$
    \item Constant bias $(1)$
    \item Video with bias $(z_v \otimes 1)$
    \item Audio with bias $(1 \otimes z_a)$
    \item Bias-bias interaction $(1 \otimes 1)$
\end{enumerate}
This approach disentangles unimodal, bimodal, and constant factors, allowing the model to learn which interactions are most informative for fraud detection.

\subsection{Fraud detection network}

The output of the Tensor Fusion layer is flattened and fed into a Fraud Detection Network consisting of:
\begin{enumerate}
    \item \textbf{Input}: Flattened tensor fusion output ($2,145$ dimensions)
    \item \textbf{Hidden layers}:
    \begin{itemize}
        \item First dense layer: $2,145 \rightarrow 128$ with ReLU activation
        \item Dropout $(0.2)$ for regularization
        \item Second dense layer: $128 \rightarrow 128$ with ReLU activation
        \item Dropout $(0.2)$ for regularization
    \end{itemize}
    \item \textbf{Output layer}:
    \begin{itemize}
        \item Final dense layer: $128\rightarrow1$ with sigmoid activation
        \item Outputs probability of fraudulent activity
    \end{itemize}
\end{enumerate}

\subsection{Model training and optimization}
The model is trained using binary cross-entropy \cite{hastie2009elements} as the loss function, with the AdamW optimizer \cite{kingma2014adam}. This setup is effective for binary classification tasks, such as fraud detection, where the goal is to minimize the difference between predicted and actual labels.
\begin{enumerate}
    \item The model was trained end-to-end using binary cross-entropy loss with the AdamW optimizer and a learning rate of $1\times10^{-4}$.
    \item Training utilized a batch size of 8 with mixed precision to optimize computational efficiency
    \item A CosineAnnealingLR scheduler \cite{loshchilov2016sgdr} adjusts the learning rate based on validation performance.
    \item Early stopping is implemented to prevent overfitting, with model checkpointing to save the best-performing model.
\end{enumerate}
The architecture is implemented in PyTorch, leveraging pre-trained weights for both ViViT and AST models to benefit from transfer learning.

\label{approach}

\section{Results and Discussion}
In this section, we present a comprehensive evaluation of our tensor fusion-based multimodal fraud detection system and compare it against baseline approaches.

\subsection{Experimental setup}

We evaluated our multimodal fraud detection model using ViViT for video feature extraction and AST for audio feature extraction. The experiments were conducted with the following configuration:
\begin{itemize}
    \item Learning rate: \(1 \times 10^{-4}\)
    \item Batch size: 4
    \item Optimizer: AdamW
    \item Scheduler: CosineAnnealingLR
    \item Training strategy: Mixed precision training with gradient accumulation
    \item Hardware: NVIDIA L40S GPU with 48GB memory ~\cite{nvidiaL40S2023}
\end{itemize}

\subsection{Dataset and evaluation metrics}

Our custom dataset consisted of 820 pairs of video-audio samples from public transportation scenarios in Rwanda, manually labeled ``Fraud'' (356 samples) or ``Legit'' (464 samples). We used a stratified 5-fold cross-validation approach to ensure robust evaluation.\\
Performance was assessed using standard classification metrics: accuracy, precision, recall, and F1 score, with particular emphasis on recall given the importance of detecting fraudulent activities.

\subsection{Performance comparison}
Table~\ref{tab:comparative} presents the comparative results of our tensor fusion approach against several baseline methods, including unimodal approaches and traditional fusion techniques.

Our tensor fusion approach significantly outperforms all baseline methods across all metrics. Compared to early fusion (simple concatenation), tensor fusion achieves a 4.9\% improvement in accuracy (89.5\% vs. 84.6\%), 4.9\% in precision (87.2\% vs. 82.3\%), 8.8\% in recall (84.0\% vs. 75.2\%), and a 7.0\% improvement in F1 score (85.6\% vs. 78.6\%). The substantial gain in recall is particularly important for fraud detection applications, as it indicates fewer missed fraud cases.\\\\
The high precision of our model (87.2\%) demonstrates its ability to minimize false alarms, which is crucial to maintaining the trust of riders in automated systems. Furthermore, the strong recall rate (84.0\%) ensures that most fraudulent activities are detected, while the resulting F1 score of 85.6\% reflects the balanced performance of our approach in the context of fraud detection.

\begin{table}[!t]
    \centering
    \caption{Performance Comparison of Fraud Detection Models}
    \begin{tabular}{|l|c|c|c|c|}
        \hline
        \textbf{Model} & \rotatebox{90}{Accuracy (\%)} & \rotatebox{90}{Precision (\%)} & \rotatebox{90}{Rec. (\%)} & \rotatebox{90}{F1 (\%)} \\
        \hline
        Video Only (ViViT) & 79.8 & 76.1 & 68.4 & 72.0 \\
        \hline
        Audio Only (AST) & 75.3 & 71.5 & 64.3 & 67.7 \\
        \hline
        Early Fusion & 84.6 & 82.3 & 75.2 & 78.6 \\
        \hline
        Late Fusion & 83.0 & 80.5 & 73.6 & 76.9 \\
        \hline
        \textbf{Tensor Fusion (Ours)} & \textbf{89.5} & \textbf{87.2} & \textbf{84.0} & \textbf{85.6} \\
        \hline
    \end{tabular}
    \label{tab:comparative}
\end{table}

\subsection{Ablation studies}
To further analyze the contribution of different components and interactions in our model, we conducted ablation studies as shown in Table~\ref{tab:ablation}. 
Ablation studies reveal several important insights:

\begin{table}[]
    \centering
    \caption{Ablation Study Results}
    \begin{tabular}{|l|c|c|c|}
        \hline
        \textbf{Model Configuration} & \rotatebox{90}{Accuracy (\%)} & \rotatebox{90}{Recall (\%)} & \rotatebox{90}{F1 (\%)} \\
        \hline
        Video Only & 79.8 & 68.4 & 72.0 \\
        \hline
        Audio Only & 75.3 & 64.3 & 67.7 \\
        \hline
        Early Fusion without Embed. & 82.5 & 72.1 & 75.4 \\ 
        \hline
        Early Fusion with Embed. & 84.6 & 75.2 & 78.6 \\
        \hline
        TF - Unimodal Only & 85.7 & 76.4 & 79.8 \\ 
        \hline
        TF - Bimodal Only & 87.8 & 80.6 & 83.1 \\ 
        \hline
        \textbf{Complete TF} & \textbf{89.5} & \textbf{84.0} & \textbf{85.6} \\
        \hline
    \end{tabular}
    \label{tab:ablation}
\end{table}

\begin{enumerate}
    \item Although video signals provide stronger fraud cues (F1 score of 72.0\%) than audio (67.7\%), both modalities capture complementary information essential for effective fraud detection.
    \item The dedicated modality-specific embedding networks before fusion improve performance by 3.2\% F1 score compared to direct feature concatenation (78.6\% vs. 75.4\%). This highlights the importance of transforming raw modality features into a suitable representation space before fusion.
    \item Bimodal interactions capture significant cross-modal dynamics, contributing to a substantial performance gain (3.3\% F1 improvement over unimodal-only tensor fusion). This validates our hypothesis that explicit modeling of modality interactions is crucial for effective fraud detection.
\end{enumerate}

The complete tensor fusion approach, which incorporates both unimodal and bimodal interactions, achieves the best performance with an F1 score of 85 6\%, demonstrating the importance of modeling all types of interactions in multimodal fraud detection.

\subsection{Error analysis}
\begin{figure}[t]
    \centering
    \includegraphics[width=0.83\linewidth]{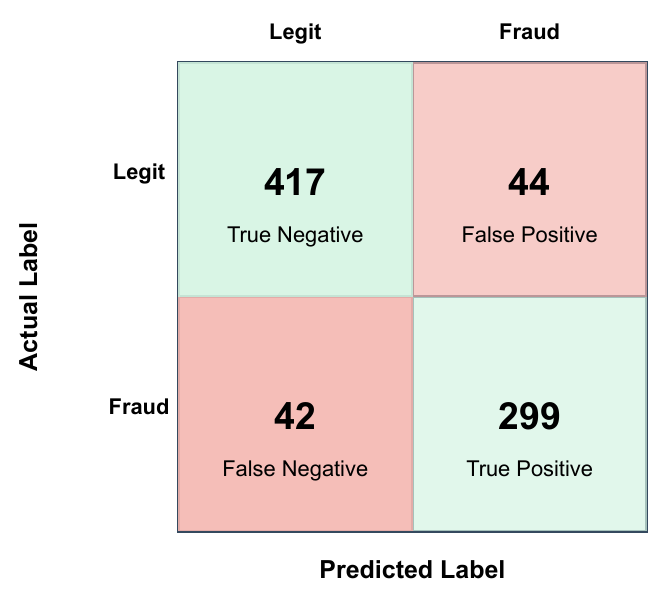}
    \caption{Confusion matrix for the Tensor Fusion model.}
    \label{fig:confusion_mat}
\end{figure}

Fig. \ref{fig:confusion_mat} presents a confusion matrix of the predictions of the tensor fusion model. The matrix illustrates the classification performance for fraud detection with 299 correctly identified fraudulent transactions (True Positives) and 417 correctly identified legitimate transactions (True Negatives). The model shows a balanced error distribution with only 44 False Positives (legitimate transactions incorrectly flagged as fraud) and 42 False Negatives (missed fraud cases), resulting in an accuracy of 89.5\%, precision of 87.2\%, recall of 84.0\%, and F1 score of 85.6\%.

The confusion matrix reveals that false negatives (missed fraud cases) occur less frequently than in the baseline approaches.\\
Qualitative analysis shows that the model particularly excels at detecting subtle fraud patterns where audio-visual correlations are important, such as distinguishing between legitimate card taps and fraudulent behaviors where passengers mimic the tapping motion without actually using their cards.

Common error cases include
\begin{itemize}
    \item Scenarios with severe visual occlusion where the payment area is not clearly visible.
    \item Instances with overwhelming background noise that masks transaction sounds.
    \item Novel fraud techniques that are not represented in the training data.
\end{itemize}

\subsection{Computational efficiency}
Despite the increased complexity of the modeling, our tensor fusion approach remains computationally efficient. The inference time on an NVIDIA L40S GPU averages 98ms per sample, enabling real-time detection at approximately 10 frames per second. The model requires 156 MB of memory, which makes it deployable on edge devices in public transportation environments.

The superior performance of our tensor fusion approach can be attributed to its ability to explicitly model both independent modality-specific patterns and their multiplicative interactions~\cite{zadeh2017tensor,li2021low,varshneya2024interpretable}. This modeling has been shown to be particularly valuable in fraud detection scenarios, where deception indicators often manifest as subtle inconsistencies between visual behaviors and audio cues~\cite{heinrich1998deception,jaiswal2019multimodal,tian2023unsupervised,tan2020detecting,wang2024multimodal}.

\label{res-disc}

\section{Conclusion}
In this paper, we present a novel approach to the detection of fraud and fare evasion in public transportation systems using a Tensor Fusion Network (TFN) that effectively combines video and audio modalities.
First, we demonstrate that explicitly modeling the interactions between visual and audio modalities through a 2-fold Cartesian product significantly outperforms traditional fusion approaches. Our tensor fusion model achieved 89.5\% accuracy, 87.2\% precision, and 84.0\% recall, representing a substantial improvement over early fusion baselines (7.0\% gain in F1 score).

Second, our ablation studies revealed the importance of modeling unimodal and bimodal interactions, with bimodal interactions providing a 3. 3\% improvement in the F1 score over unimodal-only approaches. This finding underscores the value of cross-modal analysis in detecting subtle fraud indicators that would be missed by single-modality systems.

The Tensor fusion approach represents a significant step forward in automated fraud detection for public transportation systems. By effectively capturing the complex relationships between visual behaviors and audio cues, our model provides transportation authorities with a powerful tool to reduce revenue losses, improve operational efficiency, and ensure fairness for all passengers.

\subsection{Future directions and recommendations}
Real-time processing capabilities remain a critical challenge, requiring advances in both hardware and software infrastructure. As Miller et al. \cite{miller_advancements_2024} noted, optimized algorithms and specialized hardware accelerators could enable real-time CCTV analysis for immediate fraud detection and response. Additionally, leveraging advanced machine learning techniques such as Generative Adversarial Networks (GANs) could improve anomaly detection accuracy by generating synthetic fraud scenarios for training.

Effective deployment requires strong collaboration between public transport authorities, technology providers, and legal experts. Sedmak \cite{sedmak_what_nodate} emphasized that successful implementation depends on coordinated stakeholder engagement throughout the development and deployment process. Comprehensive policies addressing ethical and privacy concerns must be established, particularly given the sensitive nature of surveillance data in public spaces \cite{li_ethical_2023}.

Technical enhancements should focus on incorporating additional contextual data such as passenger profiles and historical travel patterns, which could significantly improve fraud detection accuracy. The system must also address robustness challenges including visual occlusions, varying lighting conditions, and background noise interference. Finally, developing lightweight model architectures suitable for edge deployment would enable cost-effective scaling across entire transportation networks without requiring centralized processing infrastructure.

\label{concl}

\bibliographystyle{unsrtnat}
\bibliography{main}

\begin{IEEEbiography}
[{\includegraphics[width=1in,height=1.25in,clip,keepaspectratio]{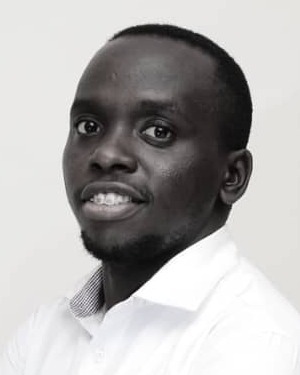}}]{Peter Wauyo} received the B.Sc. degree in computer science from Makerere University, Kampala, Uganda, in 2020, and the M.S. degree in information technology from Carnegie Mellon University Africa, Kigali, Rwanda, in 2025. He has over five years of experience as a software developer and AI engineer, having worked internationally on production-grade AI systems in domains spanning transportation, finance, and language technologies. His research interests include multimodal machine learning, natural language processing, computer vision, and intelligent transport systems. Mr. Wauyo is passionate about leveraging AI to solve real-world problems in emerging markets. 

\end{IEEEbiography}

\begin{IEEEbiography}[{\includegraphics[width=1in,height=1.25in,clip,keepaspectratio]{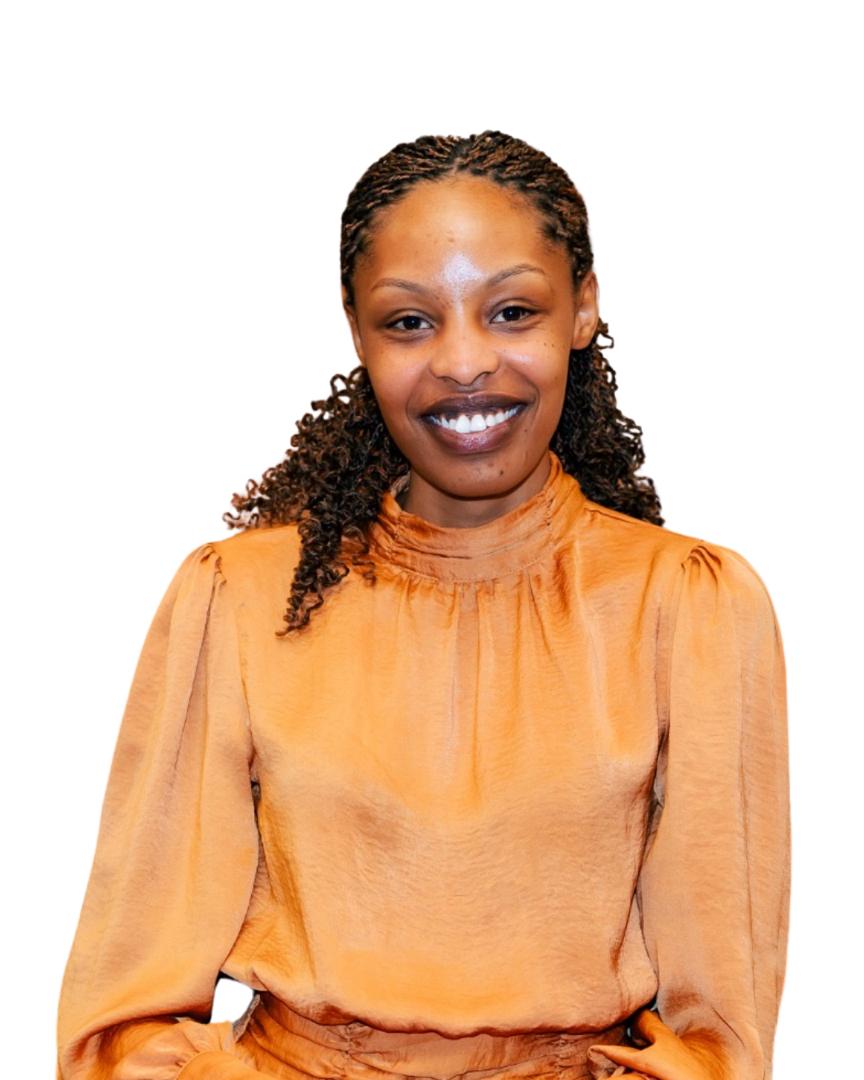}}]{Dalia Bwiza} received the B.Sc. degree in information systems from the University of Rwanda, Kigali, Rwanda, in 2022, and the M.Sc. degree in information technology from Carnegie Mellon University Africa, Kigali, Rwanda, in 2025. She has worked in machine learning and software development across sectors such as telecommunications, agriculture, finance, and education, focusing on AI-based solutions and data-driven decision making. Ms. Bwiza was awarded for academic excellence by the First Lady of Rwanda under the Imbuto Foundation initiative. Her research interests include machine learning, multimodal data fusion, intelligent transport systems, and agricultural analytics. 
\end{IEEEbiography}

\begin{IEEEbiography}[{\includegraphics[width=1in,height=1.25in,clip,keepaspectratio]{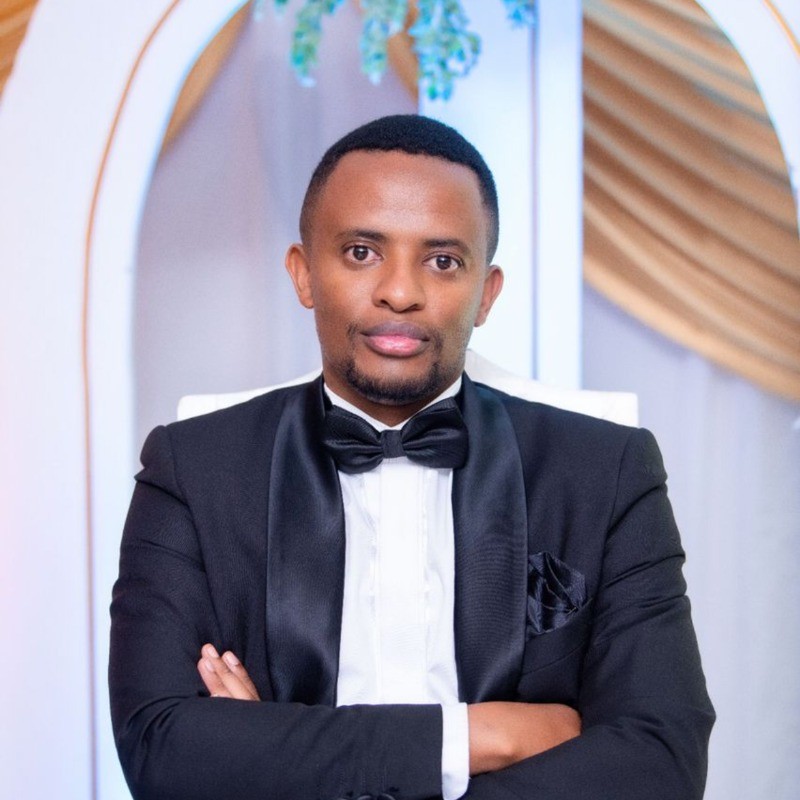}}]{Alain Murara} is a seasoned data science professional with nearly a decade of experience at the Rwanda Utilities Regulatory Authority (RURA), where he currently serves as Division Manager in Charge of Data Science and Analytics. Since joining RURA in 2016, he has held several key roles, including Senior Data Scientist, Senior Manager of Data Analytics and Knowledge Management, and Big Data Analyst. Throughout his career, Alain has led strategic initiatives focused on data-driven decision-making, regulatory intelligence, and digital transformation within Rwanda’s utilities sector.
He holds a Master’s Degree in Information Technology from Carnegie Mellon University Africa (2014–2016) and a Bachelor of Science in Computer Science and Systems from the National University of Rwanda, where he graduated with Second Class Honours, Upper Division.
\end{IEEEbiography}
\begin{IEEEbiography}[{\includegraphics[width=1in,height=1.35in,clip,keepaspectratio]{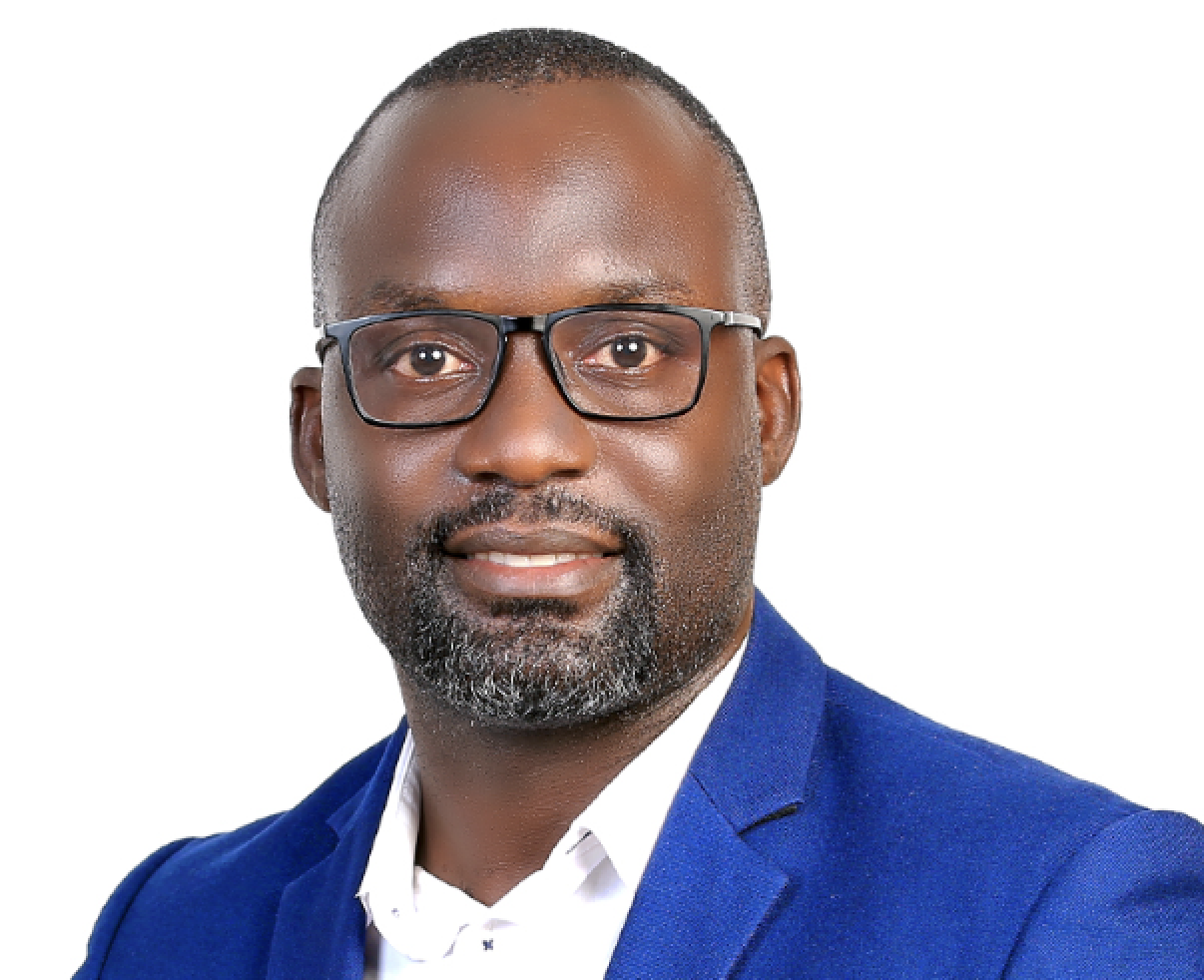}}]{Edwin Mugume} (S'13-M'17) is an Assistant Teaching Professor at Carnegie Mellon University Africa. He received the BSc degree in Electrical Engineering (First Class Hons.) from Makerere University, Uganda in 2007 and the MSc degree in Communication Engineering (with Distinction) from The University of Manchester, UK in 2011. He completed his Ph.D. in Electrical and Electronic Engineering from The University of Manchester in 2016. His Ph.D. focused on energy efficient deployment strategies for future highly dense heterogeneous cellular networks. His research interests lie in developing deployment strategies for green heterogeneous cellular networks, 5G network technologies,  Internet of Things design and applications, and machine learning applications. 
\end{IEEEbiography}

\begin{IEEEbiography}
[{\includegraphics[width=1in,height=1.25in,clip,keepaspectratio]{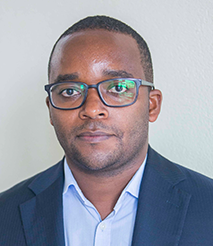}}]{Eric Umuhoza}  is an Assistant Teaching Professor at Carnegie Mellon University Africa. Before joining CMU-Africa, he held various academic and research positions across European institutions. These include serving as a senior postdoctoral researcher at the Department of Information Engineering, Computer Science and Mathematics at the University of L’Aquila (Italy), a postdoctoral researcher and teaching assistant at the Department of Electronics, Informatics and Bioengineering at the Polytechnic University of Milan (Italy), and a visiting scholar at École des Mines de Nantes (France).
Dr. Umuhoza's research interests lie at the intersection of technology and society. His work focuses on Big Data Analysis for smart applications, User Interaction Design, and Model-Driven Software Engineering. He is also committed to advancing equity and safety in public transportation, as well as promoting digital accessibility and inclusion—ensuring that digital systems and infrastructures are designed to be inclusive and equitable for all users. 
He holds a Ph.D. in Information Technology and Engineering, a Master of Science in Engineering of Computing Systems, and a Bachelor of Science in Computer Engineering, all from the Polytechnic University of Milan.
\end{IEEEbiography}






\end{document}